\documentclass[letters,fleqn,usenatbib]{mnras}

\usepackage[T1]{fontenc}
\usepackage{ae,aecompl}
\usepackage{graphicx}	
\usepackage{amsmath}	
\usepackage{amssymb}	

\newcommand{\be}{\begin{equation}}
\newcommand{\ee}{\end{equation}}
\newcommand{\bq}{\begin{eqnarray}}
\newcommand{\eq}{\end{eqnarray}}

\title[Cosmological impact of redshift drift measurements]{Cosmological impact of redshift drift measurements}

\author[J. Esteves et al.]{
J. Esteves,$^{1,2}$\thanks{E-mail: joshua.esteves.mail@gmail.com (JE)}
C. J. A. P. Martins,$^{2,3}$\thanks{E-mail: Carlos.Martins@astro.up.pt (CJAPM)}
B. G. Pereira$^{2,3,4}$\thanks{E-mail: up201705220@edu.fc.up.pt (BGP)} and C. S. Alves$^{2,5}$\thanks{E-mail: catarina.alves.18@ucl.ac.uk (CSA)}
\\
$^{1}$Université de Montpellier - 163 rue Auguste Broussonnet - 34090 Montpellier\\
$^{2}$Centro de Astrof\'{\i}sica da Universidade do Porto, Rua das Estrelas, 4150-762 Porto, Portugal\\
$^{3}$Instituto de Astrof\'{\i}sica e Ci\^encias do Espa\c co, CAUP, Rua das Estrelas, 4150-762 Porto, Portugal\\
$^{4}$Faculdade de Ci\^encias, Universidade do Porto, Rua do Campo Alegre, 4150-007 Porto, Portugal\\
$^{2}$Department of Physics and Astronomy, University College London, Gower Street, London WC1E 6BT, United Kingdom
}

\date{Accepted XXX. Received YYY; in original form ZZZ}

\pubyear{2021}

\begin{document}
\label{firstpage}
\pagerange{\pageref{firstpage}--\pageref{lastpage}}
\maketitle

\begin{abstract}
The redshift drift is a model-independent probe of fundamental cosmology, but choosing a fiducial model one can also use it to constrain the model parameters. We compare the constraining power of redshift drift measurements by the Extremely Large Telescope (ELT), as studied by Liske {\it et al.} (2008), with that of two recently proposed alternatives: the cosmic accelerometer of Eikenberry {\it et al.} (2020), and the differential redshift drift of Cooke (2020). We find that the cosmic accelerometer with a 6-year baseline leads to weaker constraints than those of the ELT (by $60\%$), but with identical time baselines it outperforms the ELT by up to a factor of 6. The differential redshift drift always performs worse that the standard approach if the goal is to constrain the matter density, but it can perform significantly better than it if the goal is to constrain the dark energy equation of state. Our results show that accurately measuring the redshift drift and using these measurements to constrain cosmological parameters are different merit functions: an experiment optimized for one of them will not be optimal for the other. These non-trivial trade-offs must be kept in mind as next generation instruments enter their final design and construction phases.
\end{abstract}



\section{Introduction}

The consequences of the fact that the redshift of objects following the cosmological expansion changes with time---the redshift drift---were first coherently formulated by \citet{Sandage,Mcvittie}, and the first detailed observational  feasibility study, for the Extremely Large Telescope (ELT), was done by \citet{Liske}. This measurement is a key science and design driver for the ELT-HIRES spectrograph \citep{HIRES}.

A redshift drift measurement is unique in comparing different past light cones (in a fully model-independent way), rather than mapping our present-day past light cone. Nevertheless, these measurements also constrain specific cosmological models. An overview can be found in \citet{Quercellini}, and more recent explorations include \citet{Kim,Second,Alves}. These show that although ELT redshift drift measurements, on their own, lead to constraints that are not tighter than those achievable from other data, they do probe different regions of parameter space, enabling the breaking of key degeneracies upon combination.

Two other approaches to detecting the redshift drift have been recently proposed. The Cosmic Accelerometer \citep{Eikenberry} aims to lower the cost of the experiment, while the Acceleration Programme \citep{Cooke} proposes to measure the differential redshift drift between two non-zero redshifts, rather than measuring the drift with respect to today. We use standard Fisher Matrix techniques in a comparative study of the cosmological impact of the three approaches. We explore relevant experimental parameters to identify relevant trade-offs, using two fiducial models: canonical $\Lambda$CDM and the Chevallier-Polarski-Linder (CPL) parametrization \citep{CPL1,CPL2}. Flat models are assumed throughout.

\section{Measuring the redshift drift}

The redshift drift of an object following the cosmological expansion, in a time span $\Delta t$, is given by \citep{Sandage,Liske,Second}
\be
\frac{\Delta z}{\Delta t}=H_0 \left[1+z-E(z)\right]\,,
\ee
although the astrophysical observable is a spectroscopically measured velocity
\be\label{specvel}
\Delta v=\frac{c\Delta z}{1+z}=(cH_0\Delta t)\left[1-\frac{E(z)}{1+z}\right]\,;
\ee
here $E(z)=H(z)/H_0$ is the rescaled Hubble parameter, with $H_0$ being the Hubble constant. We will use the CPL parametrization as our most general fiducial model. Its Friedmann equation is
\be\label{cpl}
E^2(z)=\Omega_m(1+z)^3+\Omega_\phi(1+z)^{3(1+w_0+w_a)} e^{-3w_az/(1+z)}\,.
\ee
Our primary goal is to forecast the uncertainties on cosmological parameters from redshift drift measurements, using Fisher Matrix techniques detailed in \citet{Alves}.

A detailed study of redshift measurements by the ELT has been done by \citet{Liske}. The measurement relies on absorption features in the Lyman-$\alpha$ forest and metal absorption lines redwards of it. This study found that the spectroscopic velocity uncertainty is well described by
\be\label{elt}
\sigma_v=\sigma_e\left(\frac{2370}{S/N}\right)\sqrt{\frac{30}{N_{QSO}}}\left(\frac{1+z_{QSO}}{5}\right)^{-\lambda}\, cm/s\,,
\ee
where the last exponent is $\lambda=1.7$ up to $z=4$ and $\lambda=0.9$ for $z>4$. Consistently with this and the latest top-level requirements for the ELT-HIRES spectrograph \citep{HIRES}, we assume that each measurement has a signal to noise ratio $S/N=3000$ and uses data from $N_{QSO}=6$ quasars; as baseline values we take $\sigma_e=1.35$ and an experiment time $\Delta t=20$ years, but we also explore different choices for these two parameters. The ELT can measure the redshift drift in the approximate redshift range $2.0\leq z\leq 4.5$. The main bottleneck, in addition to the spectrograph stability which is thought not to be a limiting factor \citep{Milakovic}, is the availability of bright quasars providing the required signal to noise in reasonable amounts of telescope time. The discovery of additional bright quasars will improve the experiment feasibility \citep{Boutsia}, so our ELT analysis is likely to be conservative. 

The Cosmic Accelerometer of \citet{Eikenberry} (henceforth CAC) is a low-cost version of the experiment, relying on commercial off the shelf equipment. The proposal is part of the Astro2020 decadal survey, and to our knowledge no detailed feasibility study has yet been carried out. We assume the spectroscopic velocity uncertainty given in the proposal white paper
\be\label{cac}
\sigma_v=\sigma_c\sqrt{\frac{6}{t_{exp}}}\, cm/s\,,
\ee
with the baseline value of $\sigma_c=1.5$ (but will again explore the impact of different values of this parameter), and $t_{exp}$ is the experiment time in years. The redshift ranges probed are, in principle, similar to those for the ELT. Another advantage of this approach would be a detection of the redshift drift on a time scale of 6 years.

The Acceleration Programme of \citet{Cooke} (henceforth APR) uses the ELT as described above but proposes to measure the drift between sources at two different non-zero redshifts along the same line of sight (while in the standard approach one of these redshifts is always $z=0$). The measured quantity in this case will be
\be\label{specdiff}
\Delta v_{ir}=(cH_0\Delta t)\left[\frac{E(z_r)}{1+z_r}-\frac{E(z_i)}{1+z_i}\right]\,,
\ee
where $z_r$ and $z_i$ are the redshifts of the reference and intervening sources, and $z_i<z_r$. The rationale is that for some choices of redshifts the detected drift signal can be significantly larger than in the standard approach.

Our fiducial model for the comparison will be $\Lambda$CDM, with $\Omega_m=0.3$, $H_0=70$ km/s/Mpc, $w_0=-1$ and $w_a=0$, the latter two only apply to the CPL model. Our main comparison diagnostic is the derived uncertainty in the cosmological parameters being considered.

\section{ELT versus Cosmic Accelerometer}

We will compare the ELT and CAC assuming a single measurement of the redshift drift, for a flat $\Lambda$CDM fiducial model and with priors of $\sigma_{H,prior}=10$ km/s/Mpc and $\sigma_{m,prior}=0.1$ respectively for the Hubble constant and the matter density. We explore the impact of choices of normalization of the spectroscopic velocity uncertainties ($\sigma_e$ and $\sigma_c$, collectively denoted $\sigma_z$), of the experiment time, and of the redshift at which the measurement is made. Our metric will be the posterior matter density constraint, $\sigma_m$.

\begin{figure*}
\begin{center}
\includegraphics[width=\columnwidth,keepaspectratio]{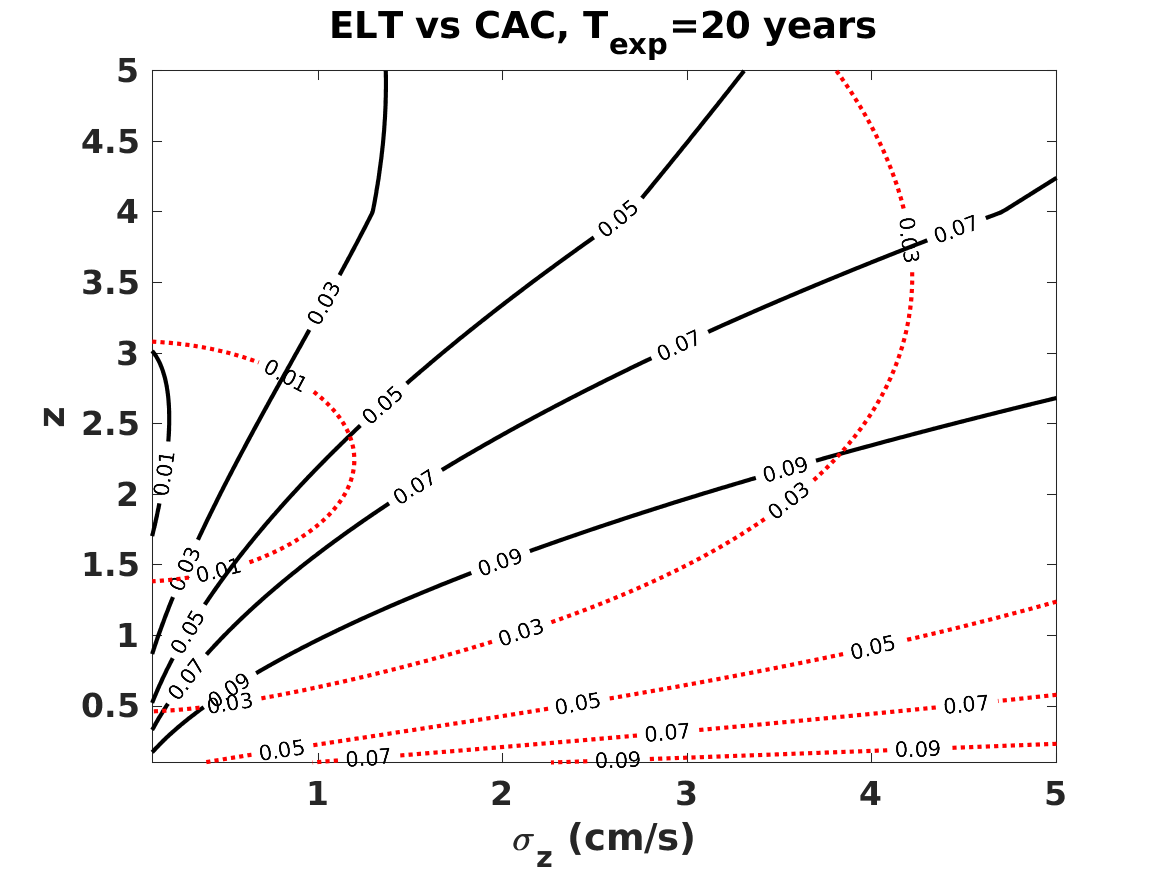}
\includegraphics[width=\columnwidth,keepaspectratio]{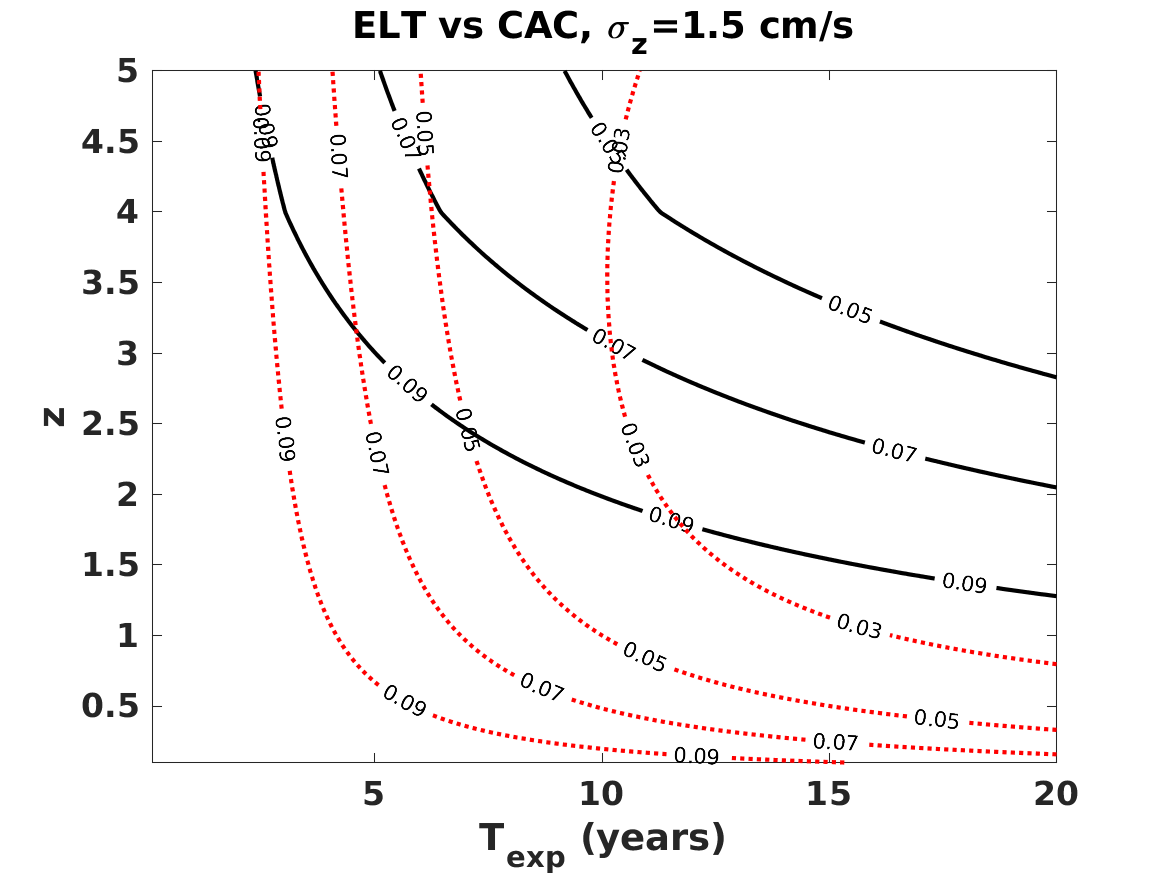}
\includegraphics[width=\columnwidth,keepaspectratio]{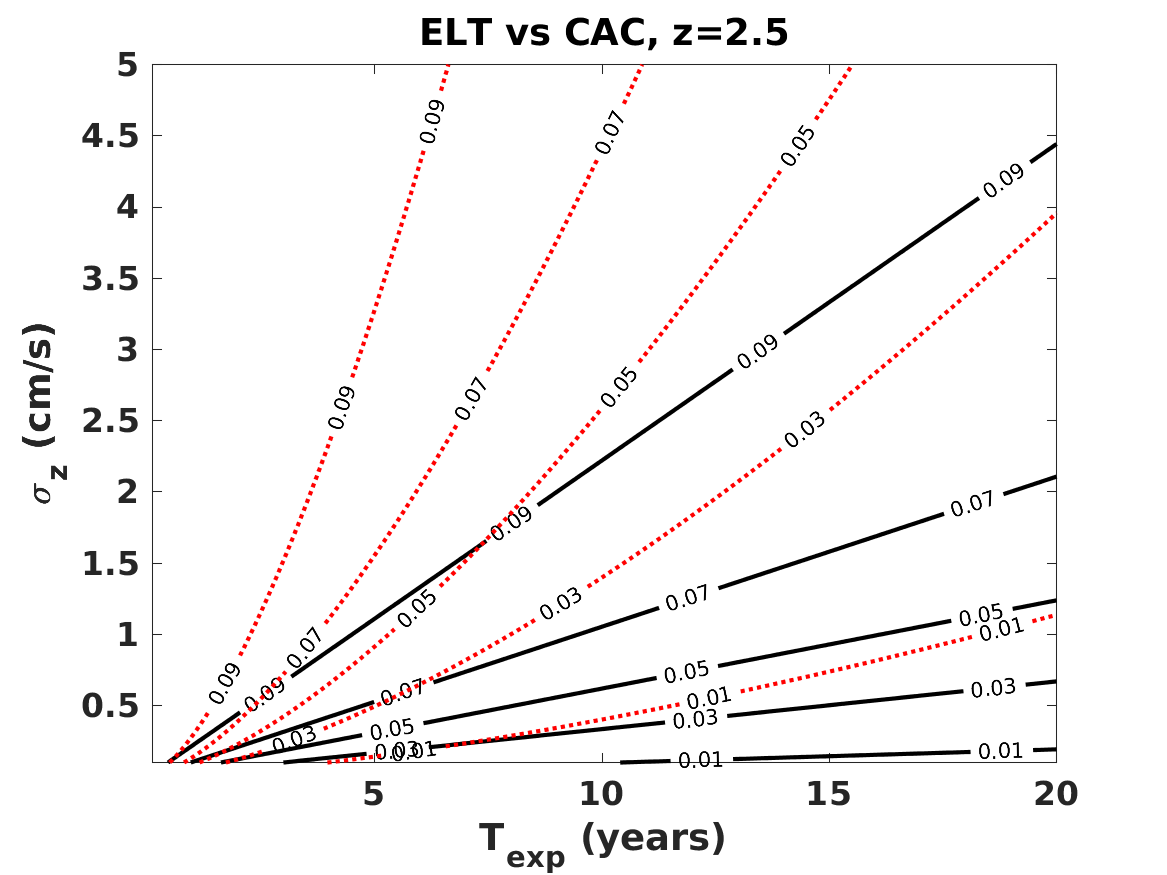}
\includegraphics[width=\columnwidth,keepaspectratio]{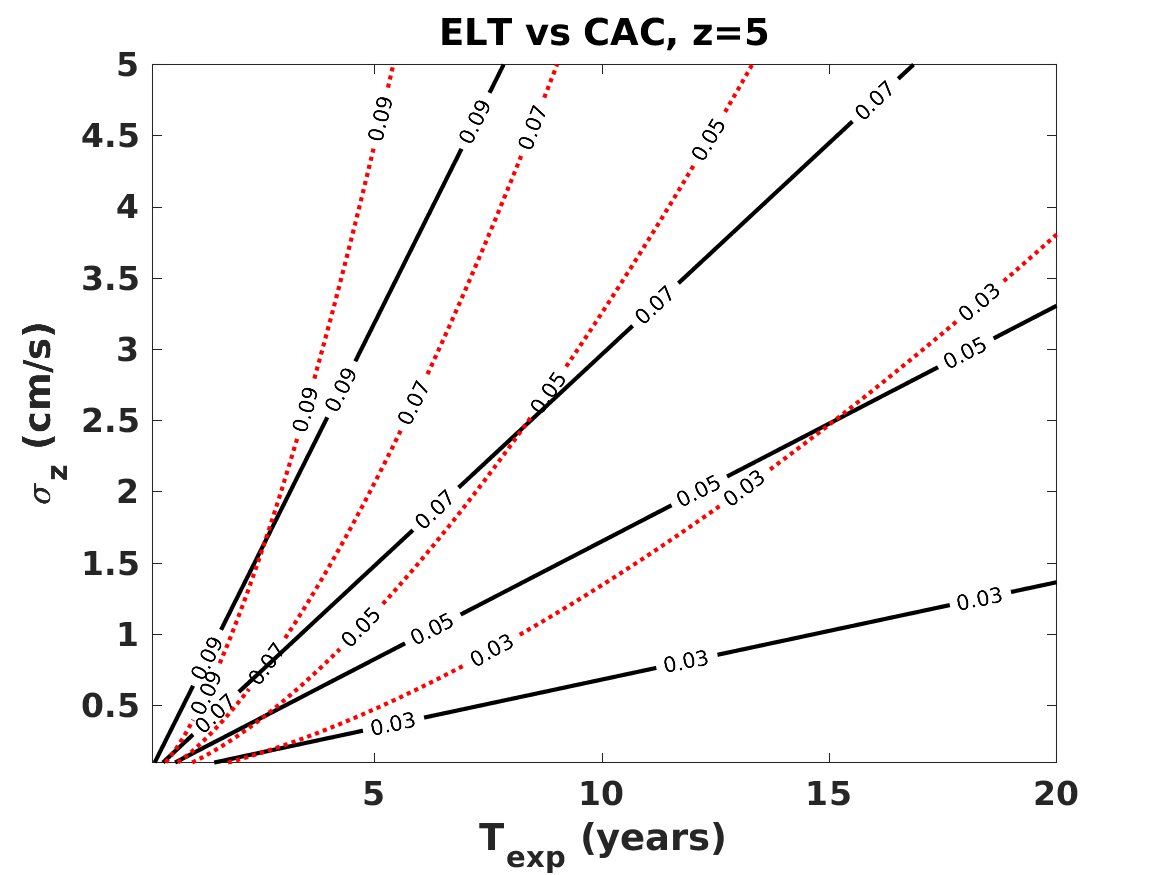}
\end{center}
\caption{Constraint on the matter density ($\sigma_m$) from a single redshift drift measurement as a function of the redshift of the measurement ($z$), experiment time ($T_{exp}$), and normalization of the spectroscopic velocity uncertainty ($\sigma_z$), assuming a prior $\sigma_{m,prior}=0.1$. Black solid and red dotted contours depict the results for the ELT and CAC experiments. For example, consider the top right panel, corresponding to measurements with ($\sigma_z=1.5$ cm/s. Then this panel shows that for one measurement at $z=3$ with an experiment time $T_{exp}=5$ years, the ELT would improve the prior constraint on the matter density to about $\sigma_{m,ELT}=0.09$ while the CAC would improve it to $\sigma_{m,CAC}=0.07$, and would therefore be more constraining.}
\label{fig1}
\end{figure*}

Importantly, cosmological parameter sensitivities of the redshift are redshift-dependent---see \citet{Alves} for a through discussion. This needs to be convolved with the spectroscopic velocity uncertainty of each experiment, bearing in mind that the ELT one is redshift dependent (cf. Eq.~\ref{elt}) while the CAC one is not (cf. Eq. \ref{cac}), at least if one takes the CAC proposal at face value.

\begin{figure*}
\begin{center}
\includegraphics[width=\columnwidth,keepaspectratio]{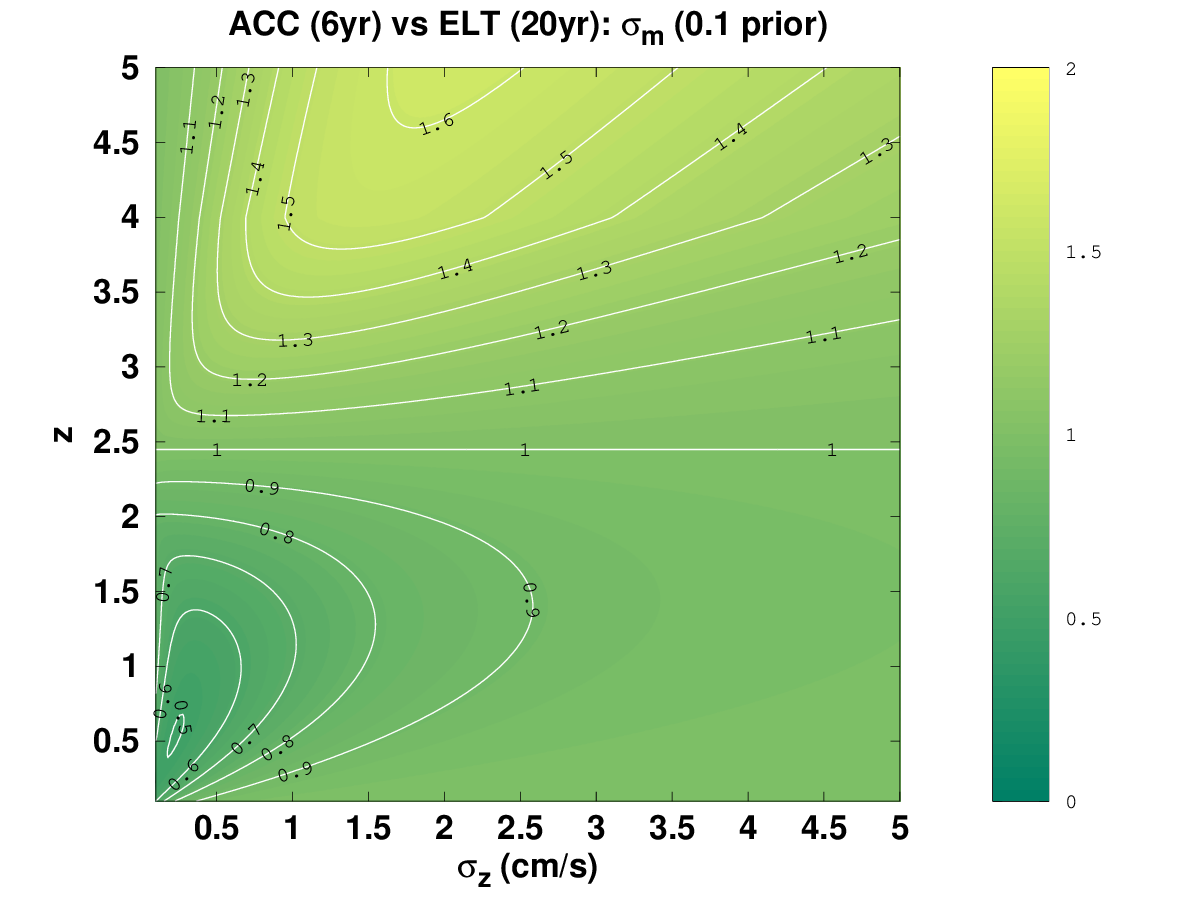}
\includegraphics[width=\columnwidth,keepaspectratio]{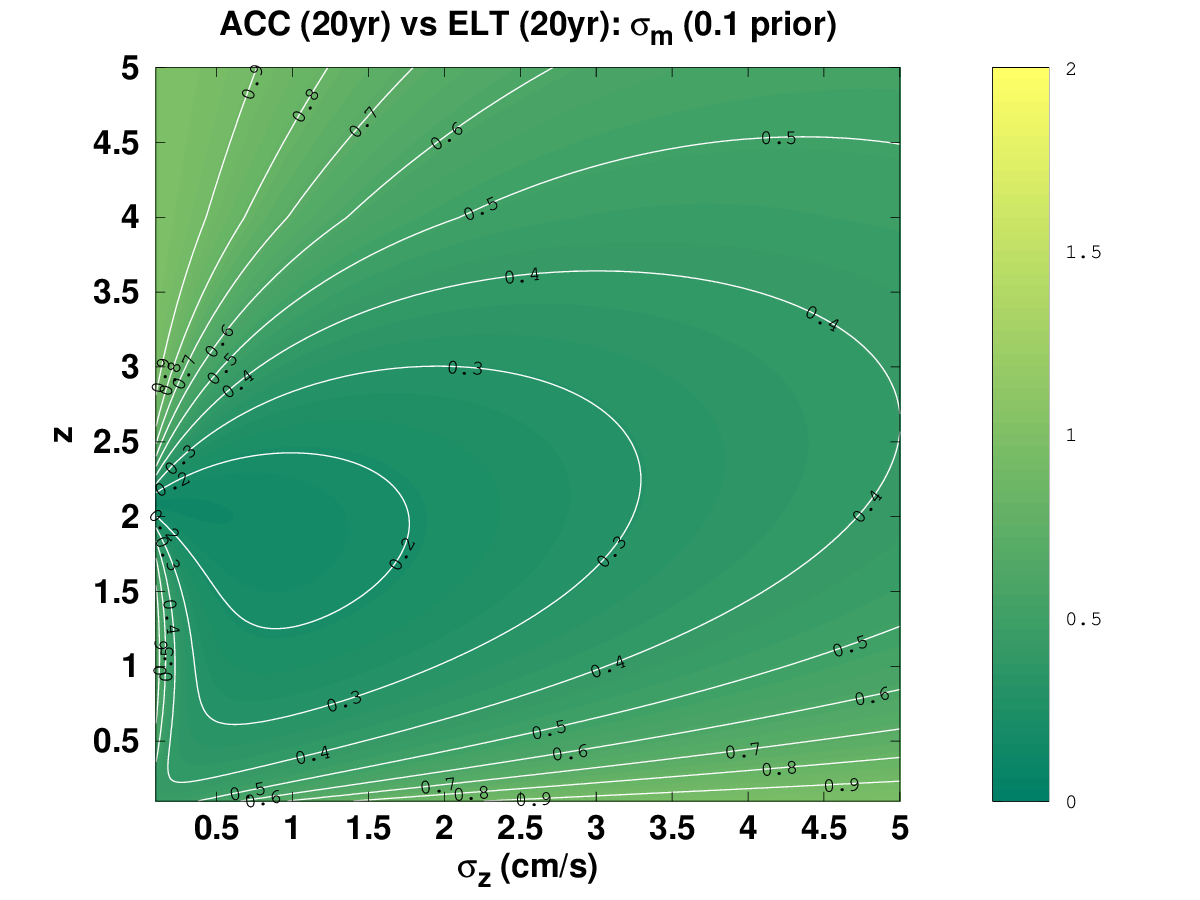}
\end{center}
\caption{Ratio of the constraints on the matter density ($\sigma_m$) from a single redshift drift measurement by the CAC and ELT experiments, as a function of the redshift of the measurement ($z$) and the normalization of the spectroscopic velocity uncertainty ($\sigma_z$), assuming an experiment time of 6 or 20 years for CAC (left and right panels respectively), and 20 years for ELT in both cases. A ratio larger than unity implies that the ELT measurement is more constraining, while a ratio smaller than unit implies that the CAC is more constraining.}
\label{fig2}
\end{figure*}

Figure \ref{fig1} shows the constraints on $\sigma_m$ for the two experiments as a function of the experiment parameters. Realistically both experiments can only do measurements at redshifts larger than $z\sim2$ (since for lower redshifts the Lyman-$\alpha$ forest falls in the ultraviolet), but we shown lower redshifts to further illustrate the different cosmological parameter sensitivities. One sees that for identical experiment times the CAC does somewhat better, though the difference decreases for larger redshifts.

A possible advantage of the CAC (in addition to cost) is a lower experiment time. The left panel of Figure \ref{fig2} compares the constraints on $\sigma_m$ for a 6 year CAC and a 20 year ELT experiment, and here the ELT does clearly better at the redhsifts of interest---up to $62\%$ better. While absolute uncertainties $\sigma_v$ may be smaller for the CAC, the relative ones will be larger due to smaller experiment times, since a key feature of the redshift drift is that the signal grows linearly in time. The right panel of the same figure compares both experiments at 20 years; in this case the CAC is more constraining---by up to a factor of six around redshift $z\sim2$. Thus an experiment optimized to detect the redshift drift signal is not necessarily optimized to constrain particular cosmological models, or indeed specific model parameters. In the example of Figure \ref{fig2} the constraints on the matter density can be substantially different, but those on the Hubble constant never differ by more than a few percent. 

\section{Canonical versus differential redshift drift}

Here we directly compare the parameter uncertainties in the canonical redshift drift and the differential redshift drift, for the CPL parametrization. To remain consistent with the work on the canonical redshift drift we will use the same five reference redshifts used by \citet{Alves}, at $z_r= \{2.0, 2.5, 3.0, 3.5, 4.5\}$, and a 20 years experiment.

In the analysis that follows we use two sets of intervening redshifts, $z_1 = \{0.67, 0.67, 0.67, 0.67, 0.67\}$ and $z_2 = \{0.0, 0.23, 0.26, 0.88,  3.0\}$. The first set of redshift is chosen to maximize the spectroscopic velocity signal: for our chosen fiducial cosmological model $z=0.67$ is the redshift with maximum positive spectroscopic velocity, and since for any redshift $z_i > 2.09$ the drift signal should be negative for that fiducial model, the difference $\Delta v_{ir}$ should be amplified and maximal. The second set of redshifts $x_2$ was numerically determined, resulting from a grid-based optimization process with the goal of maximizing the Figures of Merit for the dark energy equation of state parameters, $w_0$ and $w_a$, defined as the inverse of the area of the one-sigma confidence contour in the two-dimensional parameter space.

We note that in this optimization analysis we effectively treat these redshifts as free parameters. Admittedly this is somewhat unrealistic in that it ignores instrumental and observational limitations: one can only use the known bright quasars and its corresponding absorption systems, so regions of the experimental parameter space that would be optimal in principle may not be available in practice. Our goal in this work is to identify the conceptual advantages and disadvantages of the two methods.

For the Fisher Matrix external priors, for consistency with the earlier analysis in \citet{Alves}, we use a future priors set based on the CORE collaboration \citep{CORE}, which forecasts $\sigma_{\Omega_m h^2,CORE}=0.00028$, and on the Euclid mission, for which one expects $\sigma_{w0}=0.02$ and $\sigma_{wa}=0.1$ \citep{Euclid}.


\begin{table}
	\centering
	\caption{Ratios of the one-sigma uncertainties in the model parameters for our fiducial $\Lambda$CDM model. A ratio larger than unity favours the ELT. For convenience, the cases where APR is favoured are highlighted in boldface. The results on the left and right sides of the table correspond to the two choices of redshifts  ($z_1$ and $z_2$) described in the main text, and P refers to the priors described in the main text.}
	\label{tab:LCDM}
		\begin{tabular}{l | c c | c c}
			\hline
			Parameter & $\frac{APR1}{ELT}$ & $\frac{APR1+P}{ELT+P}$ & $\frac{APR2}{ELT}$ & $\frac{APR2+P}{ELT+P}$ \\
			\hline
			$\sigma$(h) & 1.88 & 2.63 & {\bf 0.73}  & 1.75 \\
			$\sigma$($\Omega_m$) &4.85  & 2.57 & 1.04 & 1.71 \\
		\hline
	\end{tabular}
\end{table}
\begin{table}
	\centering
	\caption{Ratios of the one-sigma uncertainties in the model parameters for our CPL model with fiducial $\Lambda$CDM model parameters. A ratio larger than unity favours the ELT. For convenience, the cases where APR is favoured are highlighted in boldface. The results on the left and right sides of the table correspond to the two choices of redshifts  ($z_1$ and $z_2$) described in the main text, and P refers to the priors described in the main text.}
	\label{tab:CPL}
		\begin{tabular}{l | c c | c c}
			\hline
			Parameter & $\frac{APR1}{ELT}$ & $\frac{APR1+P}{ELT+P}$ & $\frac{APR2}{ELT}$ & $\frac{APR2+P}{ELT+P}$ \\
			\hline
			$\sigma$(h) & 5.50  & 2.63 & {\bf 0.06}  & 1.75 \\
			$\sigma$($\Omega_m$) & 11.66  &2.57 & {\bf 0.05}  & 1.71 \\
			\hline
			$\sigma$($\omega_0$) &3.75  & 1.00 & {\bf 0.01} & 1.00 \\
			$\sigma$($\omega_a$) & 21.4 & 1.00 & {\bf 0.02}  & 1.00  \\ 
			\hline
	\end{tabular}
\end{table}


The results for the $\Lambda$CDM and CPL models are respectively listed in Table \ref{tab:LCDM} and Table \ref{tab:CPL}. These make it clear that amplifying the signal for the redshift drift does not necessarily improve the uncertainties of any parameters. However, the numerically determined set $z_2$ shows that in ideal circumstances the uncertainties for $w_0$ and $w_a$ can be improved by one to two orders of magnitude with respect to the canonical redshift drift, while the uncertainties for $h$ and $\Omega_m$ were also improved. Nevertheless this advantage is diluted when priors are added, due to relatively high non-diagonal terms in the Fisher matrices. In particular, we see that constraints on the dark energy equation of state parameters are dominated by the priors, making the ratios of the corresponding uncertainties become unity. It will be important to validate these results though a MCMC analysis, which we leave for subsequent work.

\begin{figure*}
\begin{center}
\includegraphics[width=\columnwidth,keepaspectratio]{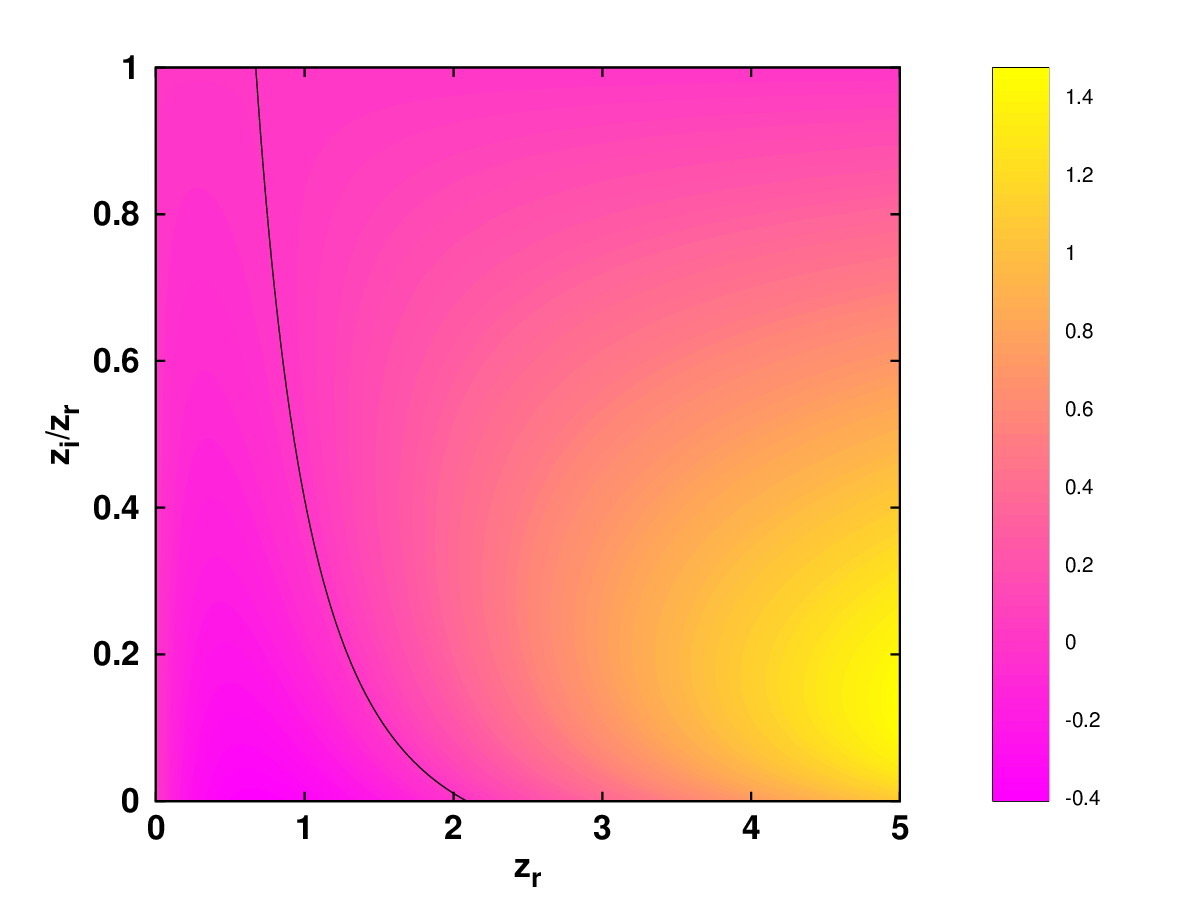}
\includegraphics[width=\columnwidth,keepaspectratio]{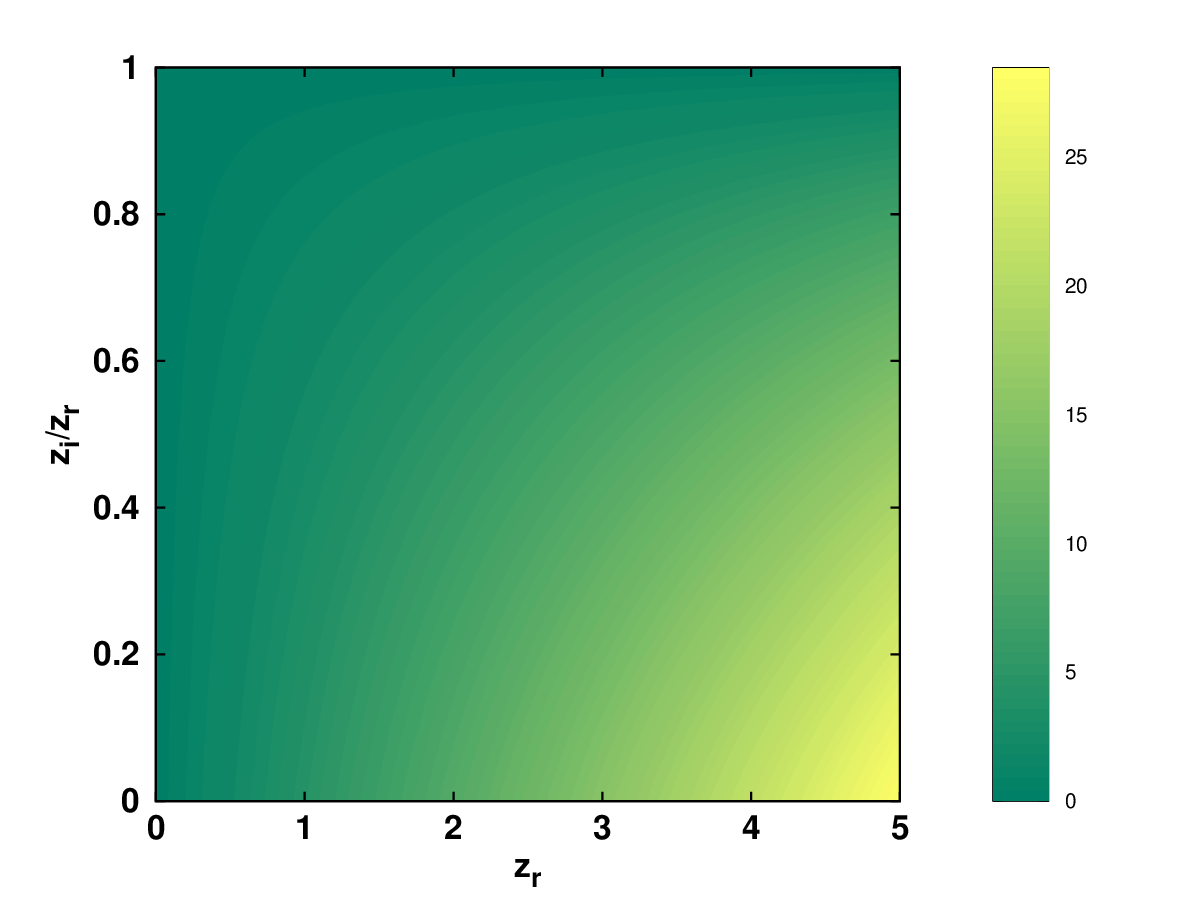}
\includegraphics[width=\columnwidth,keepaspectratio]{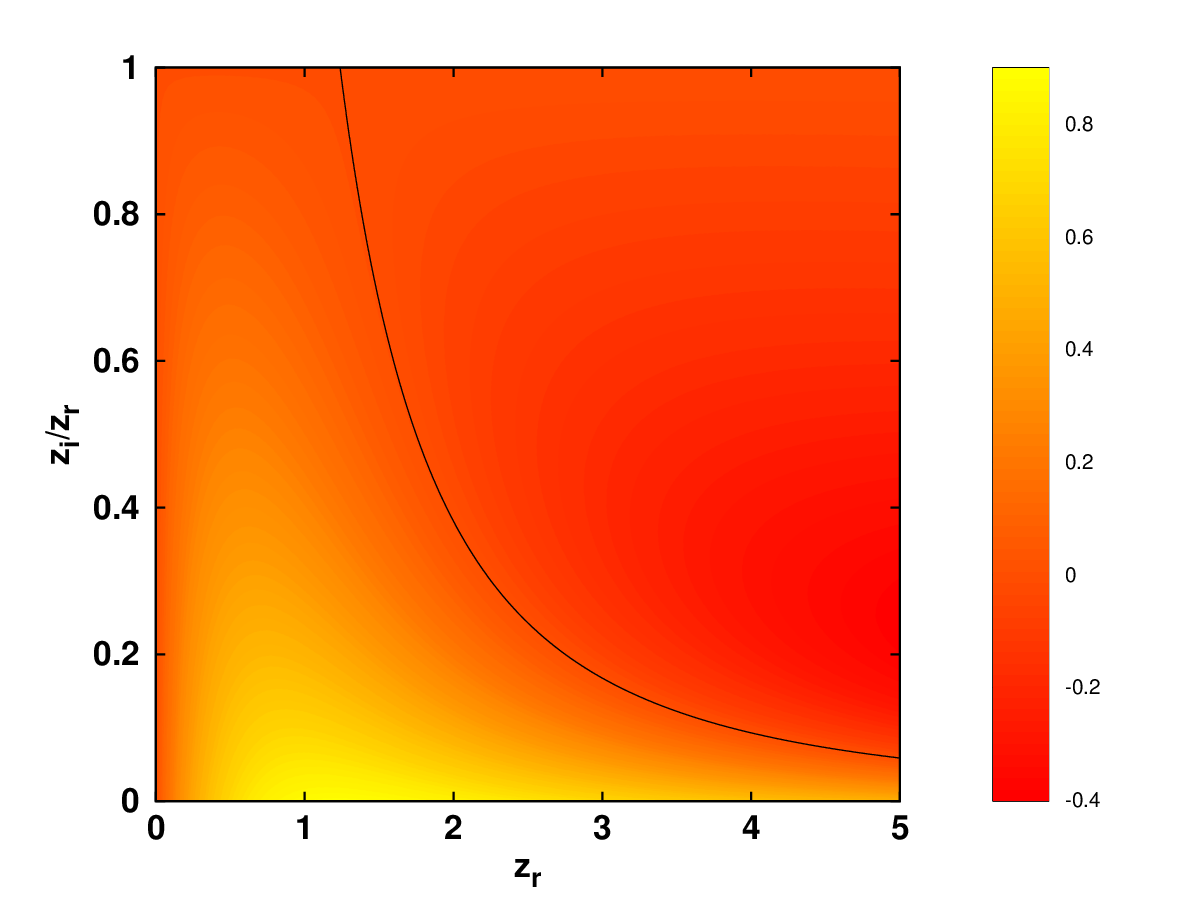}
\includegraphics[width=\columnwidth,keepaspectratio]{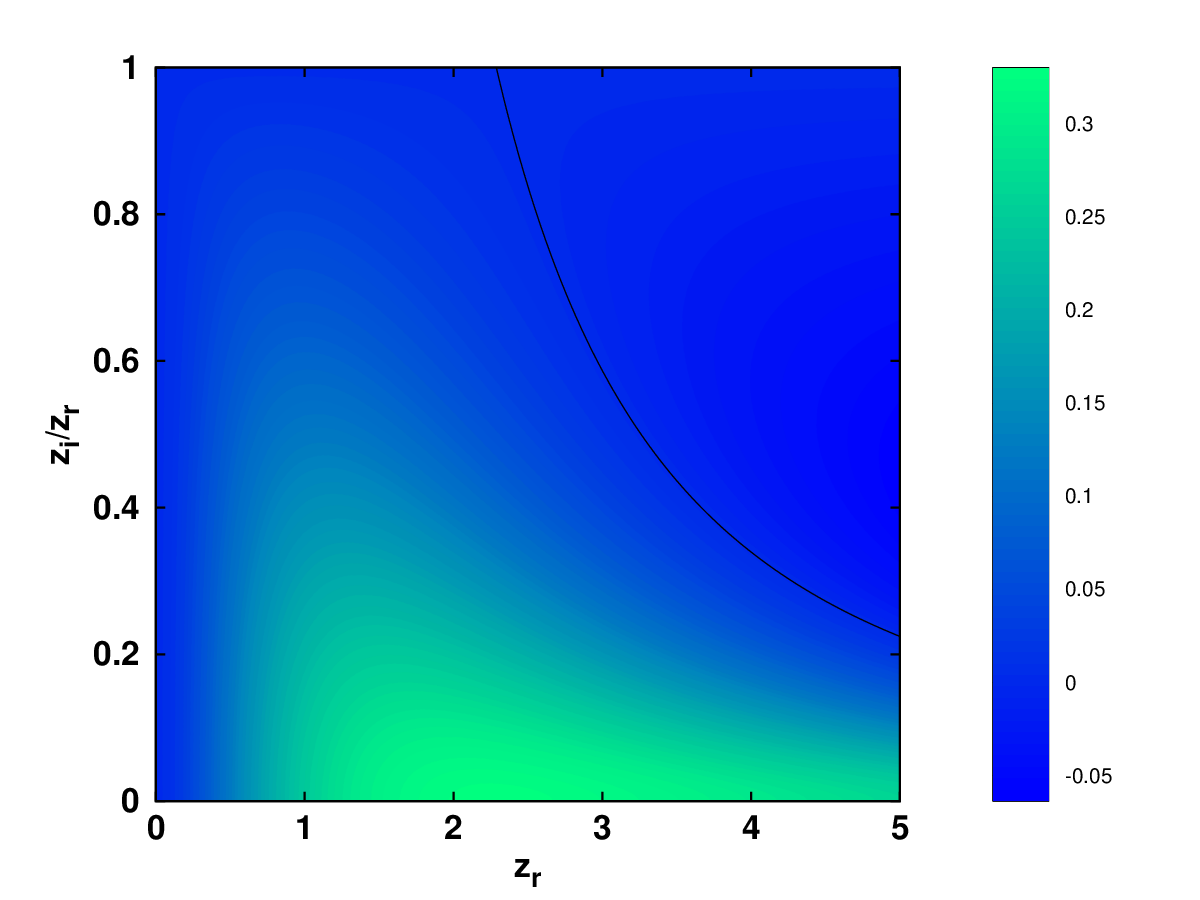}
\end{center}
\caption{The sensitivity coefficients of the differential redshift drift to the cosmological parameters $h$ (top left), $\Omega_m$ (top right), $w_0$ (bottom left) and $w_a$ (bottom right), as a function of the reference redshift $z_r$ and the ratio of intervening and reference redshifts $z_i/z_r$. The colormaps show the sensitivity in cm/s per year of observation, and the black curves show the locus of zero sensitivity. Notice that for $\Omega_m$ there is no such non-trivial locus.}
\label{fig3}
\end{figure*}

These results can be understood by calculating the sensitivity coefficients of the differential redshift drift to our four cosmological parameters, $\partial (\Delta v_{ir})/\partial p$, where $p$ is one of $(h,\Omega_m,w_0,w_a)$, which are depicted in Fig. \ref{fig3}. As expected, and already discussed in \citet{Alves}, the sensitivity to $\Omega_m$ is much larger than the others. Crucially, the sensitivity to $\Omega_m$ is maximized for $z_i=0$ and maximal $z_r$: in other words, if the goal is to constrain $\Omega_m$, the canonical redshift drift measurement is always the optimal strategy. On the other hand, for the dark energy equation of state parameters (and also for $h$) the differential redshift drift can provide tighter constraints.

These results show that an optimized observational strategy, with a specific goal, can improve the scientific outcomes. In practice, it is possible that perfect sets of reference and intervening redshifts cannot be found which will again reduce the advantage of the differential redshift drift method.

\section{Conclusions}

Our analysis shows that forthcoming redshift drift experiments need to be optimized for a specific purpose. Conceptually, such a measurement is of fundamental importance, as it will be the first time in our exploration of the distant universe that we will compare different past light cones---or, in other words, that we will see the universe expand in real time---and moreover this is done in a fully model-independent way. One might therefore optimize for the detection of the signal, in which case the differential redshift drift can be better. In this case the cosmic accelerometer can also be a viable option (given its cost advantage), provided the experiment time is large enough.

In practice, the role of the redshift drift might be that of a consistency test: standard cosmological observables lead---under some modelling assumptions---to a predicted expansion history of the universe, implying a prediction of the value of the redshift drift as a function of redshift. A direct measurement of the signal at some observationally convenient redshift will then support or rule out these modelling assumptions. The question is then at what redshift this consistency test can be done to the highest degree of statistical significance.

Finally, we can use redshift drift measurements to constrain cosmological parameters. We recall the earlier result \citep{Alves} that although on its own they yield comparatively weak constraints with respect to other probes (at least if one allows for models with many cosmological parameters), they do probe regions of parameter space that are typically different from those probed by other experiments. Moreover, this sensitivity is itself redshift-dependent. In this case one might want to optimize for these constraints, either from the redshift drift alone or in combination with other data (in this work we modelled the latter case through simple external priors). Our analysis demonstrates that the differential redshift drift always performs worse that the standard approach if the goal is to constrain the matter density, but it can perform significantly better than it if the goal is to constrain dark energy.

Our analysis assumed that measurements can be done for any choice of redshift. While conceptually this choice stems from the need to understand the differences between the various approaches and to fully explore the relevant parameter space, it is clear that observationally it is not a realistic one. The number of astrophysical targets for these measurements is small \citep{Liske,Boutsia} and typically they can only be observed from one hemisphere. This provides a strong practical limitation, and can impact the relative performance of each experiment. Indeed, we may expect that the optimal targets will depend both on the experiment and on its scientific goal. We leave the discussion of realistic observational strategies, for the known sources, for subsequent work. In any case, our results clearly show that there are non-trivial trade-offs which must be considered as next generation instruments, which will do the first redshift drift measurements, enter their final design and construction phases and their science programs are planned.

\section*{Acknowledgements}

This work was financed by FEDER---Fundo Europeu de Desenvolvimento Regional funds through the COMPETE 2020---Operational Programme for Competitiveness and Internationalisation (POCI), and by Portuguese funds through FCT - Funda\c c\~ao para a Ci\^encia e a Tecnologia in the framework of the project POCI-01-0145-FEDER-028987 and PTDC/FIS-AST/28987/2017. This work was partially enabled by funding from the UCL Cosmoparticle Initiative.

{\noindent\bf Data availability:} This work uses simulated data, generated as detailed in the text.

\bibliographystyle{mnras}
\bibliography{drift} 

\bsp	
\label{lastpage}
\end{document}